\documentclass[12pt,english]{article}
\usepackage[T1]{fontenc}
\usepackage[utf8]{inputenc}
\usepackage{lmodern}
\usepackage{natbib}
\usepackage{csquotes}
\usepackage{setspace}
\usepackage{graphicx}
\usepackage{mathrsfs}
\usepackage{amsmath}
\usepackage{amssymb}
\usepackage{babel}
\usepackage[text={16.5cm,23cm},centering]{geometry}
\expandafter\def\expandafter\quote\expandafter{\quote\small}

\makeatletter

\@ifundefined{date}{}{\date{}}
\makeatother

\begin{document}

\title{\textbf{Proposal for a degree of scientificity in cosmology}}

\author{Juliano C. S. Neves%
\thanks{nevesjcs@if.usp.br%
}}

\maketitle

\begin{center}
{\it{Centro de Ciências Naturais e Humanas, Universidade Federal do ABC,\\ Avenida dos Estados 5001, Santo André, 09210-580 São Paulo, Brazil}}
\end{center}

\vspace{0.5cm}

\begin{abstract}
In spite of successful tests, the standard cosmological model, the $\Lambda$CDM model, possesses the most problematic concept: the initial singularity, also known as the big bang. In this paper---by adopting the Kantian difference between to think of an object and to cognize an object---it is proposed a degree of scientificity using fuzzy sets. Thus, the notion of initial singularity will not be conceived of as a scientific issue because it does not belong to the fuzzy set of what is known. Indeed, the problematic concept of singularity is some sort of what Kant called the noumenon, but science, on the other hand, is constructed in the phenomenon. By applying the fuzzy degree of scientificity in cosmological models, one concludes that cosmologies with a contraction phase before the current expansion phase are potentially more scientific than the standard model. At the end of this article, it is shown that Kant's first antinomy of pure reason indicates a limit to our cosmological models.  
\end{abstract}


\section{Introduction}
The standard cosmological model, the $\Lambda$CDM model\footnote{$\Lambda$ is the cosmological constant, and CDM means cold dark matter.} or the big bang model, comes from three main ingredients: (i) Einstein's theory of general relativity, (ii) the cosmological principle, and (iii) matter-energy is described by perfect fluids at large scales. General relativity provides the field equations of gravitation, which describe the space-time fabric, the cosmological principle is the \textit{belief} according to which all typical freely falling observers are equivalent in the universe, i.e., the same phenomena are observed or, using a Kantian jargon, the same phenomena are intuited by typical observers.\footnote{\citet[p.1]{Weinberg} defines typical freely falling observers as \enquote{those that move with the average velocity of typical galaxies in their respective neighborhoods.}} In the cosmological principle, it is adopted an interpretation according to which the universe appears as homogeneous and isotropic for those observers. And, lastly, perfect fluids playing the role of the matter-energy content is a useful simplification. There exists an extra ingredient in the standard model:\footnote{For some authors, dark energy and dark matter are extra ingredients as well. According to \citet[p. 41]{Merritt}, dark matter and dark energy \enquote{are auxiliary hypotheses that were invoked in response to observations that falsified the standard model.} The author  considers dark energy and dark matter as conventionalist stratagems and non-falsifiable alternatives in cosmology.} the inflationary mechanism proposed by \cite{Starobinsky}, \cite{Guth} and others. A short period of exponential expansion or a tiny fraction of a second in which space is exponentially stretched in the early universe is an \textit{ad hoc} and \textit{a posteriori} mechanism in order to solve crucial observational problems of the standard model.\footnote{Problems like flatness, isotropy, and homogeneity in the observable universe, i.e., inflation may generate a flat, isotropic, and homogeneous universe. Moreover, the inflationary mechanism is a powerful way to provide quantum fluctuations in the early universe, which are considered as seeds for the structure formation.} However, such an important extra ingredient presents theoretical problems \citep{Steinhardt}, and its indirect observation is problematic as well. From the philosophical point of view, in turn, according to \cite{Steinhardt2}, the inflationary mechanism is a stratagem that leads to non-falsifiable statements. Assuming falsifiability as a criterion of demarcation, if the inflationary mechanism is interpreted as an \textit{ad hoc} hypothesis,  it will be a \enquote{way of evading falsification} \citep[p. 19]{Popper}.  \textit{Ad hoc} hypotheses may be a sort of Achilles' heel for Popper's falsifiability.  

On the other hand, from an observational perspective, the standard model  is grounded on three important observations: (i) abundance of both deuterium and helium isotopes, generated in the primordial nucleosynthesis, (ii) the cosmic microwave background (CMB), and (iii) the galaxies redshift. From the observational point of view, the standard model is a successful cosmological model a few seconds/minutes after the supposed big bang \citep{PlanckColl}. Assuming a hot and dense initial period of expansion (and the data indicate that), the big bang model provides the primordial nucleosynthesis and the CMB as products of such a period. The oldest data in the recent expansion phase comes from approximately a few seconds to 20 minutes after the supposed big bang during the primordial nucleosynthesis \citep{Coc,Weinberg}. CMB, in turn, appeared approximately 400 000 years after the supposed big bang \citep{PlanckColl,Weinberg}. That is, there are no data from the origin of the universe or from a singular event, or as Kant would say, there is no empirical intuition from an unconditioned event. Above all, we will never obtain a byte of information from the big bang, for science and our empirical intuition work from the empirical data, i.e., from conditioned events. Science is possible just from the empirical data. Without data, we do not have science but thoughts. We will see the Kantian difference between to think of and to cognize an object in Section 2. 

From the theoretical point of view, assuming some conditions for matter and energy, the big bang cosmology (or the Friedmann equations constructed from the Einstein equations) indicates a singular state for the universe at $t=0$, also known as initial singularity. In such a hypothetical state, physical quantities or observables would be unbounded. Geometrical quantities, like space-time curvature, would assume unlimited values as well. Moreover, assuming geometrical conditions for space-time and physical conditions for matter and energy, the singularity theorems indicate an indirect and elegant form to speak of singularities in general relativity by means of the concept of geodesic (see  \citealt[chapter 8]{Hawking}, and  \citealt[chapter 9]{Wald} for detailed studies on the singularity theorems). According to the theorems, a space-time will be singular if it has at least one incomplete geodesic. However, every theorem has conditions, and the singularity theorems assume, as I said, energy and geometrical conditions. If a cosmological model in the general relativity context does not satisfy some conditions, it will be able to avoid the big bang. Another form to avoid the big bang problem is assuming new physics near the singularity. Hypothetical theories beyond general relativity, like string theory and loop quantum gravity, and effective models like brane worlds or modified gravity also provide cosmological models without a big bang \citep{Novello}. Therefore, I will consider here two types of solutions for the initial singularity, whether from \enquote{fundamental} alternatives or from effective models: emergent cosmologies and bouncing cosmologies. In emergent cosmologies \citep{Bag} the \enquote{big bang} appears as the beginning of the recent expansion phase, and before $t=0$ it is assumed, for example, the Einstein static universe. Accordingly, our universe emerges from a different space-time or geometry, a static space-time. On the other hand, bouncing cosmologies are proposals in order to eliminate the big bang such that the conception of initial singularity is replaced by a transition between a contraction phase and an expansion phase \citep{Novello,Brandenberger}. Those two classes of alternatives will be compared to the standard model using a degree of scientificity provided by fuzzy sets and Kantian philosophy.   

The main thesis in this article is: \textit{the more data related to what is thought of, the more scientific an interpretation is}. I will adopt in some degree a Kantian definition of science. That is, before Karl Popper and his famous and controversial concept of falsifiability as a criterion of demarcation between science and metaphysical speculation,\footnote{Popper's philosophy within a Kantian or neo-Kantian tradition is found, for example, in \cite{Naraniecki}.} Immanuel Kant emphasized that empirical data is a necessary condition in order to promote thoughts to cognition (or scientific knowledge). Thus, according to scholars \citep{Gava,Janiak,Berg,Watkins,Werkmeister}, Kant may be considered as a philosopher of science as well, and, for his philosophy of science, cognition is constructed from concepts and empirical data. As I said, the big bang, as the initial singularity, does not provide data. A hot and dense initial phase of expansion only suggests (equivocally) a big bang. But the problem is even worse: according to the singularity theorems, the conception of initial singularity (or any singularity in general relativity) does not provide a concept of itself. Such as the problematic concept of noumenon in Kantian philosophy, a singularity is absent of the sphere of the understanding, thus we cannot think of a singularity or the noumenon. Therefore, a careful cosmologist should never speak of the big bang as a scientific explanation for the origin of the world. Because he/she does not possess (and will not) empirical data and a concept from the initial singularity. Insofar as, following Kant, one can distinguish to think of an object and to cognize an object, I propose a degree of scientificity from that distinction. Then cosmologies with a contraction phase and without an initial singularity will be potentially more scientific, i.e., they will provide more predictions (thought objects) and, in turn, more cognition, if such data are observed/detected.  

A degree of scientificity for objects and theories may be constructed by means of fuzzy sets. Fuzzy sets, a creation of Lotfi \cite{Zadeh}, have been applied in several areas in science and technology.\footnote{An introduction to fuzzy sets and its applications is found in \cite{Zimmermann1,Zimmermann2}.} The main feature of fuzziness is the possibility of characterizing process, systems, and events with the aid of a continuum function, the membership function, which describes the continuum degree of membership of the studied system to a specific fuzzy set. Thus, within a fuzzy approach, responses are not given only in terms of \enquote{0} or \enquote{1}, \enquote{yes} or \enquote{no}, and \enquote{true} or \enquote{false}. There are degrees of membership to a given set. As we will see, the set of what is known will be represented by a fuzzy set, and the degree of belonging of an object to such a fuzzy set will indicate its degree of scientificity. With a continuum degree of belonging to fuzzy sets, scientificity will not be conceived of from the scientific-nonscientific dichotomy. 

The structure of this paper is as follows: Sec. 2 discusses the difference between to think of an object and to cognize an object in Kant. Sec. 3 introduces the notion of the big bang (physically and geometrically); By using Kantian concepts and fuzzy sets, it is suggested a fuzzy degree of scientificity in Sec 4. Sec. 5 applies the degree of scientificity and shows how bouncing cosmologies are potentially more scientific from the future data; Sec. 6 presents the first antinomy of pure reason of Kant and indicates limits in our cosmology. Sec. 7 is devoted to the final remarks. I adopt throughout this article geometrical units, in which the speed of light in vacuum $c$ and the gravitational constant $G$ are set equal to 1.            

\section{To think of an object and to cognize an object}
In this section, thought objects and known objects will be defined in order to provide a degree of scientificity in Sec. 4. The conception or the problematic concept of the noumenon in Kant is commented and it will be related to the notion of singularity in general relativity. 

\subsection{A definition of science}
In the \textit{Critique of pure reason}, Immanuel Kant constructed an influential work that, above all, considers both science and the possibility of scientific knowledge. In the period of the \textit{Critique}, Kant's philosophy was a synthesis between empiricism and rationalism. In brief, Kant divides his main book into two parts. In the first one, which I will focus on this article, Kant studied and estimated \enquote{the building materials} (B 735)\footnote{I adopt a common convention among Kant scholars. In passages of the \textit{Critique of pure reason} \citep{Kant}, it is indicated the letter B, second edition, and the corresponding page number.} for knowledge. Such a part is subdivided into \textit{The transcendental aesthetic}, which interprets both space and time as sensible forms or pure intuitions, and \textit{The transcendental logic}, which works with the human understanding and its rules. For Kant, in the first subdivision, space and time are not considered as things in themselves, are human forms for the empirical intuition. The world, also called phenomenal world or, shortly, phenomena, is constructed by means of space and time. The chaotic stimuli, which come from outside, are modeled by our sensible forms. Space and time, as sensible forms or pure intuitions, prepare the external stimuli, giving them the way to be linked to the understanding. In the second subdivision, \textit{The transcendental logic}, Kant focused on the understanding and its \textit{a priori} rules and concepts to think of an object. Every category---like the notions of quantity, quality, modality, and relation, in which the law of causality is based on---is a concept of the understanding, which is \enquote{the faculty for thinking of objects} (B 75), that is to say, the understanding provides thought objects in a broad sense. Without the understanding, the exterior chaotic stimuli, modeled by space and time, are unthinkable. Without the empirical intuition, or the data, objects and concepts are just thoughts, and we do not have any knowledge or cognition:
\begin{quote}
Our cognition arises from two fundamental sources in the mind, the first of which [intuition] is the reception of representations (the receptivity of impressions), the second [the understanding] the faculty for cognizing an object by means of these representations (spontaneity of concepts); through the former an object is given to us, through the latter it is thought in relation to that representation (as a mere determination of the mind). Intuition and concepts therefore constitute the elements of all our cognition, so that neither concepts without intuition corresponding to them in some way nor intuition without concepts can yield a cognition. Both are either pure or empirical. Empirical, if sensation (which presupposes the actual presence of the object) is contained therein; but pure if no sensation is mixed into the representation (B 74).
\end{quote}

Independently of the nature of space and time and the origin of the understanding's \textit{a priori} concepts in Kant,\footnote{An interesting discussion on the concepts of space and time in Kant, comparing them to the relativity's
 point of view, is found in \cite{Dorato}. According to the author, it is possible to support a Kantian interpretation
  of space and time (or space-time) in the Einsteinian context.} the main issue in this article is the very difference between to think of (\textit{denken}) an object and to cognize (\textit{erkennen}) an object. Kant writes:  
\begin{quote}
To think of an object and to cognize an object are thus not the same. For two components belong to cognition: first, the concept, through which an object is thought at all (the category), and second, the intuition, through which it is given; for if an intuition corresponding to the concept could not be given at all, then it would be a thought as far as its form is concerned, but without any object, and by its means no cognition of anything at all would be possible, since, as far as I would know, nothing would be given nor could be given to which my thought could be applied (B 146).
\end{quote}
Then, without the intuition, which gives an object, a concept is not considered cognition or knowledge. Even mathematics, according to Kant, which presents \textit{a priori} objects, depends on the pure intuition to be a cognition. That is, according to the German philosopher, mathematics possesses \textit{a priori} truths and its objects are given in an \textit{a priori} or pure intuition.  In regard to physics or cosmology, the pure or \textit{a priori} intuition and the pure or \textit{a priori} concepts (categories) alongside the empirical intuition, or the data, are essential in order to provide cognition: 
\begin{quote}
The pure concepts of the understanding, consequently, even if they are applied to \textit{a priori} intuitions (as in mathematics), provide cognition only insofar as these \textit{a priori} intuitions, and by means of them also the concepts of the understanding, can be applied to empirical intuitions (B 147). 
\end{quote}
That is, in physics concepts or objects constructed by the human understanding need empirical intuitions to be a cognition. Empirical intuitions are modeled by pure intuitions (space and time) and give rise to sensible intuitions. From sensible intuitions and concepts from the understanding, according to Kant, we have the right way to know nature and its laws through physics.   

However, physics is not thought of only in terms of empirical (or sensible) intuitions, and understanding's rules and concepts. Physics, according to Kant, is a proprer science. As such it is defined in terms of systematicity, apodictic certainty, and objective grounding.\footnote{See \cite{Berg}, and \cite{Watkins} for discussions on the notion of proper science in Kant.} Systematicity ensures unity of cognition in a specific science, apodictic certainty indicates an \textit{a priori} truth, which is conceived of as necessary and universal cognition (B 4). And objective grounding illustrates that, following \citet[p. 11]{Berg}, \enquote{any proper science must be systematically ordered and constitute an interconnection of \textit{grounds and consequences}.} In modern physics, and consequently in the current cosmology, grounds mean quantum and classical fields. Every effect or consequence in the phenomenal world is produced by fields in our today's view.\footnote{I suggest that the notion of field in modern physics plays the role of the notion of substance or substratum in Kant's philosophy. According to the philosopher of Königsberg, \enquote{the substratum of everything real, i.e., everything that belongs to the existence of things, is substance, of which everything that belongs to existence can be thought only as a determination} (B 225). A classical or quantum field cannot be \enquote{observed,} only its determinations or states (particles).} With respect to apodictic certainty, we may criticize Kantian notion using advances that came out after Kant's work. For example, non-Euclidean geometries showed that old \enquote{truths} were not necessary. And in the \textit{Critique}, Euclidean or mathematical truths are considered as apodictic and \textit{a priori} certainties: \enquote{mathematical propositions are always \textit{a priori} judgments and are never empirical, because they carry necessity with them, which cannot be derived from experience} (B 14). Therefore, among the three ingredients of a proper science, only systematicity and objective grounding are tenable. Both systematicity and objective grounding provide a notion of scientificity, which will be used in order to propose a degree of scientificity by using fuzzy sets. 

But what is the meaning of data, which provide objective grounding? For Kant, as I mentioned, data means empirical intuition, which is modeled by pure intuition (space and time). The means of scientific observations in the period of the \textit{Critique} were different from today's observations.  In modern physics---in the sense of a science produced from the 20th century---data come from sophisticated machines, like LHC (Large Hadron Collider) and LIGO (The Laser Interferometer Gravitational-Wave Observatory), are captured by sensible sensors and interpreted by means of statistical models. In this sense, data are interpreted, modeled, and, above all, physical phenomena are far away from the human senses in advanced experiments. It is not exaggerated to assume a kind of Platonic \enquote{demiurgy} in modern physics today, a collective demiurgy made by thousand of researchers interpreting data in different frequencies of the electromagnetic and gravitational waves spectrum in collaborations around the world.\footnote{In \textit{Timaeus}, \cite{Plato} describes the creation of the universe using a metaphor. The maker of the universe, Demiurge, acts through thought or ideal objects as patterns in order to construct the organized world, our cosmos, from chaos (see 29a).} A large number of thought objects (mathematical and statistical objects)  is used in the data interpretation to provide phenomena representation. Thus, a known phenomenon, predicted by general relativity like the gravitational waves, is cognition after sophisticated data interpretations \citep{Abbott}. Theories and models are present in the experiment, i.e., \enquote{theory dominates the experimental work from its initial planning up to the finishing touches in the laboratory,} according to \citet[p. 90]{Popper}.  

Keeping in mind the very difference among to think of, to intuit (empirically), and to cognize an object or phenomenon, I propose to quantify the degree of scientificity of cosmological models using fuzzy sets. Science in this article is not conceived of strictly in terms of Kant's philosophy. As we saw, for Kant, a proper science like physics provides apodictic certainty. Nevertheless, today such a belief is not reasonable. Scientific results and truths after Kant's work indicate temporal and historical dependencies. As I mentioned, non-Euclidean geometries show that Kantinan apodictic certainty is not viable as something totally necessary. Then, in this article, science and a degree of scientificity are defined in terms of what is thought of and intuited (observed or measured) but assuming historical, cultural, and, above all, linguistic constraints to form that which Kant called \textit{the understanding}. In this sense, a Nietzschean view---in which the concepts of perspective and interpretation are important guides---is assumed below.\footnote{See \cite{Anderson} for an introduction to Nietzschean perspectivism and \cite{Neves1} in which it is applied to physics. \cite{Babich} presents studies on Nietzschean philosophy of science.} From Nietzschean perspectivism, truths and scientific theories are provisional, they carry strongly a dependency on language.\footnote{See \cite{Pula} for Nietzsche's relation to Sapir-Whorf hypothesis in linguistics. The Sapir-Whorf hypothesis states that the structure of a language influences the speaker's world view. In \textit{Twilight of the Idols}, \citet[p. 170, \textit{\enquote{Reason} in philosophy,} 5]{Ecce} emphasizes this point in a famous passage: \enquote{I am afraid that we have not got rid of God because we still have faith in grammar...}} For example, by considering similar philosophical systems in different countries, Friedrich \citet[p. 20, \S 20]{Beyond} says in \textit{Beyond good and evil}:
\begin{quote}
The strange family resemblance of all Indian, Greek, and German philosophizing speaks for itself clearly enough. Where there are linguistic affinities, then because of the common philosophy of grammar (I mean: due to the unconscious domination and direction through similar grammatical functions), it is obvious that everything lies ready from the very
start for a similar development and sequence of philosophical systems (...).
\end{quote}
Nietzsche emphasizes linguistic affinities, which come from an \enquote{unconscious domination} of  grammar. Even natural sciences, like physics, carry such a dependency or \enquote{domination.} In particular, regarding physics, the philosopher said in the same book: \enquote{Now it is beginning to dawn on maybe five or six brains that physics too is only an interpretation and arrangement of the world (...) and \textit{not} an explanation of the world.}\footnote{\citet[p. 15, \S 14]{Beyond}.} Such as Kant, Nietzsche assumes that the phenomenal world is not a thing in itself, which is beyond and generates the empirical world in Kantian philosophy. In this sense, \cite{Kant2} says in \textit{Prolegomena to any future metaphysics}, $\S 32$:
\begin{quote}
In fact, if we view the objects of the senses as mere appearances, as is fitting, then we thereby admit at the very same time that a thing in itself underlies them, although we are not acquainted with this thing as it may be constituted in itself, but only with its appearance, i.e., with the way in which our senses are affected by this unknown something.
\end{quote}
For Nietzsche, the phenomenal world is a product of interpretations, but in his view such interpretations depend on the human body. According to Nietzschean philosophy, Kant's faculties (like the understanding) are immanent, are body's products.\footnote{One of the clearest passages in which Nietzsche states this point of view is in \textit{Beyond good and evil}, aphorism 230: \enquote{(...) \enquote{spirit} resembles a stomach more than anything.} This point is studied in \cite{Neves1}.} Therefore, Nietzsche rejects a transcendental origin for the world, and the thing in itself, in particular, is an absurd for him:\footnote{\citet[p. 206]{Fragments}, fragment 10 [202] of 1887.}
\begin{quote}
The \enquote{thing-in-itself} absurd. If I think away all the relationships, all the \enquote{qualities}, all the \enquote{activities} of a thing, then the thing does \textit{not} remain behind: because thingness was only a \textit{fiction added} by us, out of the needs of logic, thus for the purpose of designation, communication (...).
\end{quote}
Nietzsche argues that the thing in itself is absurd because it is not even a thing. As the empirical world is not a thing in itself and knowledge depends on linguistic and corporeal issues, a natural law or a principle, according to Nietzsche, \enquote{is not a matter of fact, not a \enquote{text,} but instead only a naive humanitarian correction and a distortion of meaning that you [physicists] use in order to comfortably accommodate the democratic instincts of the modern soul!}\footnote{\citet[p. 22, \S 22]{Beyond}.} In this sense, for example, the cosmological principle, which assumes equivalence among observers in the universe, is a form of isonomy or equality before the law promoted by \enquote{democratic instincts,} as is pointed out in \cite{Neves3}. Above all, the \enquote{objective} grounding in Nietzsche's view is provided by collective interpretations and multiple perspectives. This view is adopted in this article.  

\subsection{Noumenon as absence}
For some Kantian scholars,\footnote{Nietzsche is among those who do not distinguish between thing in itself and noumenon. According to \textit{Nietzsche Source}, the word noumenon does not appear in Nietzsche' works. Thing in itself, on the contrary, is common in his books (see www.nietzschesource.org).} thing in itself is synonymous with noumenon. But this is not the case in some passages of the \textit{Critique}. According to Kant, \enquote{the concept of the noumenon is (...) not the concept of an object, but rather the problem, unavoidably connected with the limitation of our sensibility} (B 344). If it is possible a \enquote{concept} of the noumenon, it will be a problematic concept: \enquote{I call a concept problematic that contains no contradiction but that is also, as a boundary for given concepts, connected with other cognitions, the objective reality of which can in no way be cognized} (B 310). The problem regarding the \enquote{concept} of the noumenon is the absence of a specific intuition for this sort of \enquote{thing.}\footnote{In this sense, Kant indicates a hypothetical intellectual intuition (B 308).} The sensible intuition regards the phenomenon, and the \textit{a priori} intuition gives \textit{a priori} objects or concepts and, as we saw, are conditions for sensibility. In this sense, the noumenon is \enquote{a boundary concept, in order to limit the pretension of sensibility (...)} (B 111). For Al-Azm, who does not accept a total synonymy between thing in itself and noumenon, the function of the noumenon \enquote{is purely negative in that it marks the limit beyond which our concepts may not and should not go (...)} \citep[p. 519]{Al-Azm}, for the noumenon is not inside the sphere of the understanding. \cite{Al-Azm} and \cite{Palmquist} defend that the thing in itself is not identical to the noumenon because both conceptions are different perspectives of the same \enquote{thing} in Kant's philosophy: the thing in itself carries an ontological sense, and the noumenon carries an epistemological sense. Therefore, following those authors, the noumenon is the limit of science, that is to say, it is the limitation of both the understanding's categories and the empirical data. The noumenon is neither a thought object nor an intuited object, then it is not a known object. And, following Nietzsche, by considering a thing in itself as an absurd, the noumenon tends to absurd because it is a limit of rationality. As we will see, the problematic concept of the noumenon finds a new form in the \enquote{concept} of singularity in general relativity, i.e., both problematic concepts are impossible to define and to intuit because they are absent of both the understanding and the phenomenal world.

\section{The big bang as the initial singularity}
Before the definition of degree of scientificity using Kantian notions listed above, it is indicated the conception or the problematic concept of the big bang from two point of views in this section: physical and geometrical point of views. From the two point of views, it is suggested that the \enquote{concept} of the big bang is some sort of  noumenon from Kantian philosophy. 

\subsection{Physical point of view}
From a physical perspective, the big bang as the initial singularity is \textit{suggested} from the Friedmann-Lemaître-Robertson-Walker (FLRW) metric \citep{Friedmann,Lemaitre,Robertson,Walker}, which is the standard class of geometries in cosmology and is a solution of the Einstein field equations $G_{\mu\nu}=8\pi T_{\mu\nu}-\Lambda g_{\mu\nu}$, in which $G_{\mu\nu}$ is the Einstein tensor, $T_{\mu\nu}$ is the energy-momentum tensor, $g_{\mu\nu}$ is the metric tensor, and $\Lambda$ is the cosmological constant---the most acceptable candidate to be interpreted as the source of cosmic acceleration so far. The FLRW metric is written as 
\begin{equation}
ds^2 =g_{\mu\nu}dx^\mu dx^\nu =  - dt^2 +a(t)^2\left[ \frac{dr^2}{1-\mathbb{K}r^2}+ r^2 \left( d\theta^2+\sin^2\theta d\phi^2  \right) \right],
\label{FLRW} 
\end{equation}
where $a(t)$ is the scale factor, which indicates, for example, the rate of expansion of the space-time fabric, and $\mathbb{K}$ is the spatial curvature of the universe ($\mathbb{K}=0$, flat universe; $\mathbb{K}=-1$, open universe; $\mathbb{K}=1$, closed universe). According to the recent data \citep{PlanckColl}, the universe is almost flat, then $\mathbb{K}\approx0$.  

Thus, by using Einstein's equations, the FLRW metric provides the Friedmann equations, which indicate the space-time dynamics: 
\begin{equation}
\left(\frac{\ddot{a}}{a}\right)  =  -\frac{4\pi}{3}\left(\rho+3P \right),
\label{Friedmann1}
\end{equation}
and
\begin{equation}
\left(\frac{\dot{a}}{a} \right)^2  =  \frac{8\pi}{3}\rho - \frac{\mathbb{K}}{a^2}.
\label{Friedmann2}
\end{equation}
The parameters $\rho$ and $P$ are, respectively, the energy density and pressure of matter-energy in the universe, conceived of as perfect fluid(s) in the standard model as we saw in Introduction, and $\dot{a}$ and $\ddot{a}$ stand for first and second derivative of $a(t)$ with respect to time, respectively. As we can see, due to observations that indicate $\dot{a}(t)>0$, i.e., an expanding universe, one can consider an initial state in which $a=0$ when $t=0$. Then, from Eqs(\ref{Friedmann1})-(\ref{Friedmann2}), it is common to interpret $t=0$ as the big bang, where physical quantities like $\rho$ diverge or $\rho \rightarrow \infty$ at $t=0$. Besides the energy density, which is a scalar, other scalars constructed, for example, from the curvature tensor diverge at $t=0$. Divergence of scalars is commonly interpreted as an indirect evidence of singularities in space-times. However, as \citet[chapter 9]{Wald} points out, there are space-times with well-behaved scalars that possess singularities. Then the scalar divergence is not a crucial criterion in order to predict singularities in space-times in general relativity. The singularity theorems, as we will see, indicate a stronger criterion.

Accordingly, the big bang is not an explosion, something that occurred in a \enquote{place.} On the contrary, it is interpret as the beginning of space-time and all matter-energy content in our world. And for that reason, several textbooks emphasize inappropriately that the big bang is the origin of the universe. \citet[p. 99]{Wald}, for example, writes that \enquote{general relativity leads to the viewpoint that the universe began at the big bang.} In the same direction, \citet[p. 340]{Carroll} is more emphatic: \enquote{it [the big bang] represents the creation of the universe from a singular state (...).} Following Kant, I argue that such a language is not appropriate in science. The singular state or origin of the universe is not a scientific matter because it is impossible to obtain any data from the unconditioned origin. Insofar as the big bang is neither an empirical event (because there is no space-time at $t=0$) nor related to any previous empirical event, it will be considered unconditioned. Moreover, to think of the big bang as \enquote{the creation of the universe} will suppose either an author beyond space and time or \enquote{metaphysical} objects that generated the world.

However, textbook authors and researchers admit that the big bang is a breakdown or limitation of general relativity. According to them, a future quantum theory of gravity would solve and eliminate such a problem replacing general relativity at scales around the Planck length, $l_p \approx 10^{-35}$ m. But if the singular event is a breakdown of general relativity, why do they say general relativity indicates the singular origin of the universe? General relativity should be silent. Nevertheless, it is possible to build non-singular solutions of the gravitational field even in the general relativity context. On the one hand, in black hole physics, there exist regular black holes, which do not present a singularity inside the event horizon \citep{Ansoldi,Neves_Saa}. On the other hand, bouncing cosmologies and emergent cosmologies are options to the standard model, are regular cosmological models in which violations of specific conditions indicated  in the singularity theorems take place at $t\approx 0$ \citep{Novello,Brandenberger,Neves2,Bag}. As we will see, in these cosmological models beyond the $\Lambda$CDM model, there is no singular initial state for the universe.   

In another front, in the geometrical point of view, the problem of the big bang is posed in a more elegant form. With the singularity theorems, singularities inside black holes and the initial singularity are \textit{supposedly} natural results of general relativity. Independently of space-time symmetries, singularities appear in solutions of Einstein's field equations. However theorems depend on conditions. To violate the singularity theorems means to violate its conditions.   

\subsection{Geometrical point of view}
From a geometrical point of view within the general relativity context, the problem of singularities is discussed by means of the singularity theorems developed by Hawking, Penrose, Geroch, and others \citep{Hawking2,Hawking,Wald}. The main idea is to use the notion of geodesic to speak indirectly of singularities. In general relativity \citep{Einstein} one may distinguish three types of trajectories: timelike, spacelike, and null. According to the convention adopted in this article, given a tangent vector $v^\mu$ to any trajectory, the norm of $v^\mu$ is negative for timelike trajectories, positive for spacelike trajectories, and it vanishes for null trajectories. If we consider only the gravitational field, such trajectories will be called geodesics, and the vector $v^\mu$ will be parallel transported along the trajectory. The main result in the singularity theorems relates geodesic continuity to existence of singularities. For a given space-time that possesses a singularity, there will be at least one incomplete geodesic. Then, the singularity theorems indicate \textit{the end of the world} (or the beginning).  

Interestingly, such theorems have been persuasive for a large number of researches. It is not exaggerated to say the singularity theorems are the main reason for many scholars to advocate the existence of singularities in the world. Let us focus on the fourth singularity theorem in Wald's book, the theorem 9.5.4 \citep[p. 240]{Wald}:
\begin{quote}
Suppose a space-time $(M,g_{\mu\nu})$ satisfies the following four conditions. (i) $R_{\mu\nu}v^\mu v^\nu \geq 0$ for all timelike and null $v^\mu$, as will be the case if Einstein's equations is satisfied with the strong energy condition holding for matter. (ii) The timelike and null generic conditions are satisfied. (iii) No closed timelike curve exists. (iv) At least one of the following three properties holds: (a) $(M,g_{\mu\nu})$ possesses a compact achronal set without edge [i.e., $(M,g_{\mu\nu})$ is a closed universe], (b) $(M,g_{\mu\nu})$ possesses a trapped surface, or (c) there exists a point $p \in M$ such that the expansion of the future (or past)  directed null geodesics emanating from $p$ becomes negative along each geodesic in this congruence. Then $(M,g_{\mu\nu})$ must contain at least one incomplete timelike or null geodesic.  
\end{quote} 
In the above theorem, any space-time, like the FLRW indicated by the metric (\ref{FLRW}), is represented as a manifold $M$ endowed with a Lorentzian metric $g_{\mu\nu}$. The first condition may be translated into $\rho+3P \geq 0$ using the FLRW metric and the definition of the Ricci tensor $R_{\mu\nu}$, and it will be one of the most important if we want to violate the theorem. Such a condition is one of the energy conditions, it is the strong energy condition. By violating it, i.e., assuming $\rho+3P <0$ for the matter-energy content, it is possible to build nonsingular cosmologies in which the big bang is absent. And energy conditions violations are more acceptable after observation of cosmic acceleration. The second condition of the theorem says that given a singular space-time, it possesses at least one point in each geodesic (whether timelike or null) experiencing a tidal force, which is generated by space-time curvature, or the gravitational phenomenon. The third condition ensures causality protection and nonexistence of time machines (something \textit{reasonable} in a real universe). And the three possibilities for the fourth condition mean: (a) the first is not according to observations (for our universe is almost spatially flat, i.e., $\mathbb{K}\approx 0 $), but it is possible a closed universe near the supposed big bang; (b) speaks of trapped surfaces, which are related to boundaries in space-time; and the last property (c) indicates a congruence of past or future null geodesics and its convergence. The cosmic expansion indicates a past convergence at the initial singularity. Then by observing the four conditions of the theorem it is possible to conceive of our universe as singular in the past, and there exists \enquote{at least one incomplete timelike or null geodesic,} such that the notion of singularity is indirectly thought of, i.e., it is indicated by means of incomplete geodesics. 

Therefore, we have neither a definition of singularity nor a defintion of the big bang provided by the singularity theorems and its conditions. \citet[p. 216]{Wald} assumes such an impossibility and says \enquote{one does not have a completely satisfactory general notion of singularities.} In this sense, I argue that the conception of singularity (whether inside a black hole or the cosmological initial singularity) resembles the conception of the noumenon, i.e., it is a problematic concept. Kant emphasized that the concept of the noumenon is \enquote{a boundary concept, in order to limit the pretension of sensibility (...)} (B 111). It is not possible to think of the noumenon because the understanding works with objects and their unproblematic concepts. Besides, the noumenon does not provide data. As the notion of space-time is not valid at the singularity, i.e., a singularity is not \enquote{a place,} we do not have an empirical intuition or data from it. Without the understanding with its rules and concepts, a singularity is unthinkable and unknowable such as the noumenon. Without any intuition (whether pure or empirical), a singularity is not a cognition. Thus, the conception of singularity and the big bang are limitations on scientific knowledge---above all, they are absent of science.\footnote{A similar conclusion is shown in \cite{Romero}, in which space-times singularities are interpreted as nonphysical realities and defective products of Einstein's theory, since energy conditions are obeyed.} As we will see in the next section, from a degree of scientificity, it is suggested that the big bang is the worst scientific idea in today's science. 

The two point of views on singularity illustrated above only suggest a singular origin for the universe. The singularity theorems should not be read from an \enquote{ontological} point of view. For from an \enquote{ontological} point of view, the singularity theorems (with their conditions) would speak of an absurd, the big bang as a thing in itself. On the other hand, from an epistemological point of view, the singularity theorems indicate the limitation of general relativity (and science) and lead to the problematic concept of the big bang as noumenon.  By not observing the theorems' conditions, general relativity would be free of a singular event at $t=0$, however, as we will see, Kant's first antinomy of pure reason does not rule out a problematic concept of the noumenon for cosmological models without the initial singularity.    

\section{A fuzzy degree of scientificity}
In this section, I propose a notion of degree of scientificity applied in cosmology. For that reason, I will conceive of thought objects, intuited objects, and known objects as members of fuzzy sets using their definitions from Kantian philosophy. A fuzzy degree of scientificity is an alternative to a binary reasoning. Instead of \enquote{scientific} or \enquote{nonscientific} attributes, it proposes quantitative answers. Within the studied context in this article, fuzzy sets are more appropriated because we may assume degrees for what is thought of and intuited. That is, it is possible to indicate a scale in which objects are more or less thought of and intuited. Consequently, as we will see, there will be a scale for what is known and then a degree of scientificity. 

Therefore, let $\mathcal{O}$ be the set of all objects (in a broad sense), denoted generically  by $o$, i.e., $o \in \mathcal{O}$. Let $O_t$ be the set of what is thought of (objects of study indicated by means of statements\footnote{Following Nietzsche, as the thing in itself is an absurd, then every object is given by the understanding's properties and relations in statements.}) and $O_i$ be the set of intuited phenomena or observed objects, which in Kantian philosophy are given by the empirical intuition. As we saw, that which is known in natural sciences is provided by both the understanding, \enquote{the faculty for thinking of objects} (B 75), and empirical intuitions. Then the set of what is known is defined as
\begin{equation}
O_k = O_t \cap O_i.
\label{S}
\end{equation}
However, $O_k$, $O_t$, and $O_i$ are not ordinary or crisp sets. Fuzzy sets may be more appropriate in this case because they provide the notion of degree of belonging of elements or objects to sets. Thus, it is possible to quantify the presence (giving the degree of belonging) or not of an object in the three sets indicated above. In terms of fuzzy sets, $O_t$ and $O_i$ are written as
\begin{equation}
O_t = \lbrace \left(o, f_t(o) \right) :o \in \mathcal{O}, f_t(o):\mathcal{O}\rightarrow [0,1] \rbrace,
\end{equation}
and
\begin{equation}
O_i = \lbrace\left(o, f_i(o) \right) :o \in \mathcal{O}, f_i(o):\mathcal{O}\rightarrow [0,1] \rbrace.
\end{equation}
The functions $f_t(o)$ and $f_i(o)$ are the membership functions of $O_t$ and $O_i$, respectively, and give the degree of belonging to the respective sets: the closer to 0, the less thought or intuited, and, inversely, the closer to 1, the more (well)-thought or (well)-intuited an object is. In ordinary sets, the image of the membership functions assumes only 0 or 1, i.e., in crisp sets we would have $\mathcal{O} \rightarrow (0,1)$ instead of $\mathcal{O}\rightarrow [0,1]$. In ordinary sets, an object belongs or not to a given set. In fuzzy sets, on the other hand, there is a degree of belonging of an element to a given fuzzy set. 

The set $O_t$ is composed by what is predicted (to be observed as phenomenon), given by statements, non-observed objects, which appear as tools to think of, and objects conceived after observation. It is straightforward to see that $O_t > O_k$, i.e., the number of that which is thought of in any model or theory is larger than the number of that which is known. Accordingly, $O_t$ is similar to Popper's world 3, the \enquote{world} of statements and theories. In \textit{Unended quest}, \citet[p. 211]{Popper2} discusses a hypothetical world's structure, divided into three \enquote{worlds}:
\begin{quote}
If we call the world of  \enquote{things}---of physical objects---the \textit{first world}, and the world of subjective experiences (such as thought processes) the \textit{second world}, we may call the world of statements in themselves the \textit{third world}. (I now prefer to call these three worlds \enquote{world I}, \enquote{world 2}, and \enquote{world 3}...). 
\end{quote}  
Statements and, especially, basic statements are used to falsify theories in Popper's epistemology. Here thought objects, given by statements, when corroborated in the phenomenal world provide cognition. As pointed out before using Nietzschean philosophy, theories are provisional interpretations and depend on human language. Popper agrees with Nietzsche in this point, indicated in the following passages: \enquote{I regard world 3 as being essentially the product of the human mind,} more precisely, \enquote{I regard the world 3 of problems, theories, and critical arguments as one of the results of the evolution of human language} \citep[p. 217]{Popper2}. 

The data, or Kantian empirical intuition, indicated by $O_i$, do not support every phenomena predicted by theories, then a lot of thought objects will be always mere speculations, whether due to illogical thinking,\footnote{According to \citet[p. 98]{Leech}, logical and real possibilities, in Kantian philosophy, mean that \enquote{something is \textit{logically possible} just when the concept of it is non-contradictory, and something is \textit{really possible} just when the concept of it is non-contradictory \textit{and} consistent with the a priori constraints on experience arising from the forms of intuition and the categories.} In our science, the word logic does not present the same meaning compared to Kant's science. Non-classical logics, like the fuzzy logic, challenges the traditional view.} excess of imagination or insufficient experiments, which are an origin for the large distance between theoretical and experimental physics today. Objects in the Planckian scale (or beyond) like strings make predictions and theories just speculations because we do not have technology to \enquote{observe} them. But in some cases, the data or the objects of  $O_i$ precede the \textit{correct} way of thinking a phenomenon. In the case of the black body radiation, the data did not agree with a continuum form to represent the electromagnetic radiation. A new object was thought of, and \cite{Planck} introduced the concept of quantum of energy to better interpret the black body radiation.   

The degree of an object in $O_i$, indicated by $f_i(o)$, may be translated into the confidence level. According to statistical models, the best experiments in physics (for example, LHC and LIGO) offer five (or more) sigma confidence level for the data. Therefore, for $o \in O_i$, the larger the confidence level, the larger $f_i(o)$.      

Following \citet{Zadeh} and his definition of intersection of two fuzzy sets, the intersection of  the two fuzzy sets $O_t$ and $O_i$ provides the relation between their membership functions and the membership function of $O_k$, i.e.,
\begin{equation}
f_k(o) = \mathrm{Min} \left[f_t(o),f_i(o) \right] \forall o \in \mathcal{O}.
\label{Degree}
\end{equation}    
The lowest value between $f_t(o)$ and $f_i(o) $ yields the value of $f_k(o)$. Then the function $f_k(o)$ gives a quantitative notion of degree of scientificity for an object and, consequently, for a model or theory. \textit{The larger $f_k(o)$, the larger the degree of scientificity}. Given any theory/model, the values of $f_k(o)$ for all objects within such a theory/model display its fuzzy degree of scientificity.  

It is possible to relate the degree of scientificity of an object $o$ in $O_k$, indicated by $f_k(o)$, to the degree of falsifiability defined by Popper. For Popper, \enquote{the amount of empirical information conveyed by a theory, or its empirical content, increases with its degree of falsifiability} \citep[p. 96]{Popper}. Insofar as $f_k(o)$ measures both the theoretical and the empirical content of an object (given by statements), the degree of an object in $O_k$ may be applied in order to indicate the degree of falsifiability of statements. Several experiments and sigmas concerning the same object alongside  well-thought and \enquote{simpler} statements, which are \enquote{falsifiable in a higher degree,}\footnote{According to Popper, following Weyl, simpler theories or statements provide less parameters (see \citealt[p. 127-8]{Popper}). The falsifiability of an statement decreases with the number of parameters.} increase $f_k(o)$, or the degree of scientificity.  

It is appropriate to illustrate an example of $f_k(o)$, the degree of scientificity. The cosmic microwave background or CMB and the dark energy may provide a clear example of the degree of scientificity of two objects in cosmology. The CMB is an almost homogeneous radiation generated approximately 400 000 years after the supposed big bang in the period in which matter (atom nuclei plus electrons) and electromagnetic radiation were in thermodynamic equilibrium. The CMB is well-described as a black body radiation with a temperature of 2.7 K today and is well-measured by sophisticated radio telescopes. Then $f_t(CMB)$ and $f_i(CMB)$  are high in the scientificity scale. Dark energy, on the other hand, presents problems \citep{Carroll2}. Dark energy is the origin for the cosmic acceleration and is thought of as the cosmological constant in the $\Lambda$CDM model. However, such an interpretation is not the only one. It is possible to think of dark energy as vacuum energy, effect of extra dimensions, and other hardly falsifiable possibilities. There exist several alternatives to dark energy today, and a consensus is not present in physics and cosmology. That is, this confusing context or theoretical weakness implies that $f_t(DE)$ is low for dark energy (DE). But from the observational point of view, dark energy has a better status indicated by supernovas, thus $f_i(DE)>f_t(DE)$. However, by comparing CMB and dark energy, it is straightforward to say $f_t(CMB)>f_t(DE)$ or $f_i(CMB)>f_t(DE)$. Therefore, with the aid of Eq. (\ref{Degree}), one concludes that the CMB degree of scientificity is higher than the dark energy degree of scientificity (i.e., $f_k(CMB)>f_k(DE)$). The CMB phenomenon is more scientific than the dark energy phenomenon today. 

In this sense, dark matter (DM) is similar to dark energy. Dark matter has some degree in $O_i$, i.e., $f_i(DM)\neq0$ because there are indirect observations that concern it, for example, rotation of galaxies and gravitational lensing \citep{Freese}. However, such as dark energy, there is no consensus about the origin and nature of dark matter. Neutrinos, axions, and even extra dimensions effects are possibilities to think of  dark matter. Thus, dark matter presents a low degree of scientificity, and such as dark energy, it is less scientific than CMB, that is to say, $f_k(CMB)>f_k(DM)$.       

By comparing models and theories in the same field, the more data related to what is thought of, the more scientific a model/theory is. As known objects are scientific objects, a theory or model may have $n$-known objects. Considering the set of what is known as a finite set, the cardinality of $O_k$ is defined as\footnote{Definitions and operations with fuzzy sets may be found in \citet{Zimmermann1,Zimmermann2}.}  
\begin{equation}
\vert O_k \vert = \sum_{j=1}^{n}f_k(o_j)=f_k(o_1)+f_k(o_2)+...+f_k(o_n). 
\label{total}
\end{equation}
In ordinary sets, the cardinality yields the number of elements of a set. In fuzzy sets the idea of cardinality is adapted using the continuum membership function. Thus, $\vert O_k \vert$ provides the degree of scientificity of model/theory with $n$-known objects. In particular for cosmological models, as we will see, $\vert O_k \vert$ is potentially larger for models with a contraction phase. There could be data from a previous phase in our universe in order to increase $\vert O_k \vert$ and to promote cosmologies with \enquote{pre-big bang} phenomena as potentially more scientific than the standard model.\footnote{Popper defended a content measure of a theory and justified that Einstein's theory has a greater content than Newton's theory: \enquote{This makes Einstein's theory \textit{potentially} or \textit{virtually} the better theory; for even before any testing we can say: if true, it has the greater explanatory power. Furthermore, it challenges us to undertake a greater variety of tests} \cite[p. 53]{Popper3}.}

\section{Application of the degree of scientificity in cosmological models}
I will apply the fuzzy degree of scientificity, defined in Eqs. (\ref{Degree}-\ref{total}), to three different types of cosmological models and their objects: the standard model, emergent cosmologies, and bouncing cosmologies. The focus is on (potential and real) known objects. Then, models with larger number of data and larger number of predicted objects or phenomena will be favored. In this sense, bouncing cosmologies are better options to the early cosmology (and moreover options to a contraction phase, before $t=0$) and, consequently, to the standard model. 

\subsection{$\Lambda$CDM model}
As we saw, the standard model is based on three main observations and three theoretical ingredients. Moreover, it is necessary an extra ingredient in order to justify the data, which indicate an almost flat, homogeneous and isotropic universe at large scales, and small inhomogeneities in the CMB. A supposed quantum field would have stretched space-time in the early universe providing an origin for the data. Besides, quantum fluctuations in this hypothetical field, according to inflationary mechanism, would have generated seeds for the structure formation (structures like galaxies). Such a quantum field and the supposed inflationary period are \textit{ad hoc} responses to the data and to the observable universe such as we have observed or intuited. An inflationary period does not solve the big bang problem because it is geodesically incomplete \citep{Borde}\footnote{\cite{Steinhardt} and \cite{Steinhardt2} argue other problems in the inflationary mechanism like multiverse, unpredictability, and the trans-Planckian problem.}, that is, from the observational point of view, the standard model even with the inflationary artifact would work a fraction of a second after the supposed big bang.

According to the above discussion, it is possible to say that the initial singularity is theoretically ill-posed, i.e., it is not an object for the understanding, it is a problematic concept or a noumenon. In terms of thought objects in general relativity, the initial singularity or the big bang (BB) possesses $f_t(BB) = 0$ or $BB \notin O_t$. Moreover, due to the impossibility of detecting it, one has necessarily $f_i(BB)=0$, which is translated into $BB \notin O_i$ and, consequently, $BB \notin O_k$. Thus the  degree of scientificity, given by Eq. (\ref{Degree}), is zero for the initial singularity. The absence of both concept and data undermines the big bang as a scientific topic in physics. In terms of phenomena, the big bang as the origin of the world will be impossible to science because it will be unconditioned. According to Kant, the empirical world---the physical or phenomenal world---is the conditioned world. As we saw, a proper science in Kantian philosophy is defined in terms of objective groundings. Every empirical data is consequence of a cause, i.e., it is conditioned. Therefore, standard model's big bang is unconditioned and would be out of the empirical world. Without data and concept, the conception of the big bang is the worst idea in science today.    

The inflationary mechanism, in turn, is a known object of study in cosmology. There is, at least, a low degree of scientificity for inflation because it is defined in terms of a field but it is indirectly intuited. Inflation possesses several theoretical problems, as indicated in \cite{Steinhardt,Steinhardt2},\footnote{\cite{Linde} tries to answer this criticism.} and a possible direct observation of the inflationary period would be provided by cosmological gravitational waves produced during space-time stretching. As is well known, inhomogeneities in the CMB are interpreted as an indirect observation of inflation. But, as we will see, bouncing cosmologies are candidates in order to explain such data without an inflationary mechanism. With a low degree of scientificity, whether due to both theoretical and observational problems, inflation justifies the search for options that concern the early cosmology.  

According to Eq. (\ref{total}), each standard model's accomplishment increases its degree of scientificity. Each known object in the $\Lambda$CDM model improves its degree of scientificity and increases the cardinality of $O_k$, which may be written as 
\begin{equation}
\vert O_k \vert_{SM}= f_k(o_1)+f_k(o_2)+...+f_k(o_l),
\label{SM}
\end{equation}
in which all $l$ known objects $\in O_k$ for the $\Lambda$CDM model.

\subsection{Emergent cosmologies}
Following \citet{Bag}, emergent cosmologies are options to the early standard cosmology before inflation. In general, models of emergent cosmologies adopt the inflationary mechanism and try to extend our knowledge to something before the inflationary period and the supposed big bang. Before the nonscientific initial singularity, emergent cosmologies assume a special cosmological model: the Einstein static universe. Albert Einstein's static model was built in 1917 \citep{Einstein2} to conceive of a non-expanding universe. By adopting the cosmological constant and matter (described as dust) in order to provide a static world, the Einstein static universe would be eternal one. However, after Einstein's work, it was shown that such a cosmological model is unstable. Emergent cosmologies use the idea that the inflationary period was generated by a static universe that went out of equilibrium. Such as the $\Lambda$CDM and its inflationary mechanism, the emergent scenario suffers from fine-tuning problems \citep{Ellis}. It is possible to construct emergent cosmologies in contexts beyond general relativity (like modified gravity, loop quantum gravity, and brane worlds), and, according to \citet{Bag}, a stable and eternal Einstein static universe before $t=0$ is produced  in such contexts beyond general relativity. However, the fine-tuning problem remains even in that contexts beyond general relativity. On the other hand, the big bang problem---still included in the standard model's inflationary mechanism---is evaded because emergent cosmologies assume geometrical conditions that do not satisfy the singularity theorems' conditions \citep{Ellis}.   

In this class of cosmologies, it is worth listing ideas of a non-empirical \enquote{world} before the big bang. Metaphysical\footnote{Metaphysical in Aristotelian sense, or at least in the sense of commentators on Aristotle, i.e., beyond nature (\textit{physis} in Greek). Nature is empirical and as such it is thought of and intuited from space and time.} proposals like quantum foam and models that suppose unobservable objects beyond space and time (or space-time) are attempts of conceiving of our world from a transcendental origin, emerging from non-empirical objects.\footnote{See \cite{Ashtekar} for some proposals in a supposed quantum big bang.} In Kantian (and Nietzschean) sense, they will be metaphysical proposals and speculations beyond scientific possibilities because, as we saw, physics and cosmology are possible only in the phenomenal world.     

However, emergent cosmologies that adopt an Einstein static universe are potentially more scientific than the standard model because it is suggested data from the previous world. As is pointed out in Eq. (\ref{total}), the degree of scientificity depends on the number of known objects in a model/theory. According to \cite{Bag}, gravitational waves generated by the oscillating Einstein universe would be (supposedly) detectable. Thus, an emergent scenario may possess a degree of scientificity with data before the inflationary period, i.e., considering that scenario, $f_i \neq 0$ and, consequently, $f_k \neq 0$ before the inflationary period. Nevertheless, emergent scenarios with inflation present the same problems of the standard model regarding the inflationary mechanism. Moreover, a realistic emergent cosmology in agreement with the data \enquote{is possible to construct only in a small region of parameter space, and that too for a rather restrictive class of inflationary potentials} \citep[p. 2]{Bag}. Potentially, as I said, emergent cosmologies are more scientific than the $\Lambda$CDM because they extend the possibility of data to a previous world, the Einstein static universe, and match the standard model from the inflationary period. Thus, the degree of scientificity of emergent cosmologies may be potentially indicated as
\begin{equation}
\vert O_k \vert_{EC}= f_k(o_1)+f_k(o_2)+...+f_k(o_l)+f_k(o_{l+1}),
\label{EC}
\end{equation}       
where the $(l+1)th$ known object would correspond to gravitational waves emitted by the Einstein static universe. As we can see, emergent cosmologies may cover all results of the $\Lambda$CDM model and may provide even more known objects, then improving its degree of scientificity ($\vert O_k \vert _{EC}>\vert O_k \vert_{SM}$).

\subsection{Bouncing cosmologies} 
A bouncing model is not the antipode to the $\Lambda$CDM model. In general, bouncing cosmologies are alternatives to the standard model only for the early universe and before $t=0$. That is, according to \citet[p. 837]{Brandenberger}, \enquote{a bounce provides a natural extension of the usual standard model of cosmology.} Then, all data from  the primordial nucleosynthesis and thereafter are adopted in bouncing cosmologies. The three main observational accomplishments for the standard model are maintained in bouncing cosmologies because the dispute concerns cognition in the early universe and before the bounce ($t=0$). In this sense, bouncing models are able to provide a transition between a contraction and an expansion phase, and the big bang as the initial or singular state of the universe is not present.      
 
Physicists and cosmologists have created alternative cosmological models in order to avoid the big bang problem. Even in the general relativity realm, bouncing cosmologies may violate some energy conditions and provide models without the initial singularity. According to researches in cosmological models beyond the standard model, sophisticated bouncing cosmologies manage to avoid not only the big bang problem. Other standard model problems (flatness, homogeneity, horizon and isotropy problems, which  are solved by adopting the inflationary mechanism in the $\Lambda$CDM model and in emergent cosmologies) are resolved by the supposed previous phase of our universe. Then bouncing cosmologies may present solutions to the standard cosmology problems and point out to the absence of an initial singularity. In a large list of bouncing options beyond the standard model, there exist, for example, the ekpyrotic cosmology, the matter bounce model, string gas cosmology and others \citep{Brandenberger,Lehners}. Moreover, according to  \cite{Steinhardt}, and \cite{Brandenberger}, several bouncing cosmologies without the inflationary mechanism do not suffer from the multiverse-unpredictability and trans-Planckian problems, which appear in inflation. Above all, as an alternative to inflation, bouncing cosmologies must provide an origin for the CMB inhomogeneities. Models like the ekpyrotic and matter bounce generate an almost scale-invariant spectrum of fluctuations in the previous contraction phase and predict parameters like the spectral index in agreement with the latest data (such as the inflationary mechanism).

However, bouncing cosmologies go beyond, whether from either the observational point of view or the theoretical point of view. From the theoretical point of view, bouncing models would appear from quantum theory of gravity candidates, like the ekpyrotic (from string theory) and models that come from loop quantum gravity. From the observational point of view, assuming a previous phase of cosmic contraction, bouncing cosmologies amplify the potential data in relation to the standard model. Phenomena in the previous phase would be detectable or indirectly intuited: black holes \citep{Quintin,Neves4}, gravitational waves \citep{Barrau}, and, as I said, even quantum fluctuations from the contraction phase would increase the degree of scientificity of bouncing cosmologies. That is, if phenomena conceived of as phenomena from the contraction phase are intuited or observed, bouncing cosmologies (BC) will increase their degree of scientificity compared to the standard model or emergent cosmologies, i.e., $\vert O_k \vert _{BC}>\vert O_k \vert _{EC}>\vert O_k \vert _{SM}$ because 
\begin{equation}
\vert O_k \vert_{BC}= f_k(o_1)+f_k(o_2)+...+f_k(o_l)+...+f_k(o_{l+m}),
\label{BC}
\end{equation}
in which $m>0$, and the $m$ known objects would come from the contraction phase. Therefore, by comparing Eqs. (\ref{SM}), (\ref{EC}), and (\ref{BC}), bouncing cosmologies are potentially more scientific among our modern options. However, as is known, bouncing cosmologies are not trouble-free: instabilities and anisotropies during the contraction phase are open issues and challenges for this class of cosmological models \citep{Brandenberger}.

The standard model with the inflationary mechanism is limited to a fraction of a second after the supposed initial singularity and it is sentenced to this temporal limit. Bouncing cosmologies extend data possibility to a previous cosmic phase, nevertheless, as we will see, they will have also a limit suggested by Kant's first antinomy of pure reason.

\section{A commentary on the first antinomy of pure reason}
A Kantian antinomy is made of thesis and antithesis.\footnote{See \citet{Grier} for an introduction to Kant's antinomies.} The philosopher of Königsberg tried to show in the \textit{Critique of pure reason} that both thesis and antithesis of an antinomy are incompatible, and, above all, scientific knowledge is not able to adopt definitely whether the thesis or the antithesis. The reason why the thesis and the antithesis are impossible in science is we do not have an empirical intuition or observation regarding either the thesis or the antithesis. And, as we saw, according to Kant, cognition depends on the empirical intuition. In particular, the first antinomy is the cosmological antinomy,\footnote{The second antinomy talks about the world and its parts, the third is about freedom, and the fourth antinomy of pure reason concerns God.} which treats the notion of beginning (or not) of the universe. The first antinomy of pure reason in the \textit{Critique} is:  
\begin{quote}
$\bullet$ \textit{Thesis}:
The world has a beginning in time, and in space it is also enclosed in boundaries (B 454).
\end{quote}
\begin{quote}
$\bullet$ \textit{Antithesis}:
The world has no beginning and no bounds in space, but is infinite with regard to both time and space (B 455).
\end{quote}
It is possible to criticize the thesis and the antithesis using arguments that came out after Kant. Non-Euclidean geometries and, consequently, general relativity provide models of finite and non-limited universes, i.e., time may have had (or not) a beginning but space may have no boundaries in the case of positive spatial curvature in the FLRW metric ($\mathbb{K}>0$ in Eq. (\ref{FLRW})). On the other hand, Olbers' paradox would be an impossibility to the antithesis.\footnote{Olbers' paradox says an infinite and eternal universe would provide bright nights for us due to an infinite number of stars.} However, somehow Kant argues against both the thesis and the antithesis because, above all, they are point of views in which the universe is thought of as a thing in itself, and the antinomy is \enquote{a conflict due to an illusion arising from the fact that one has applied the idea of absolute totality, which is valid only as a condition of things in themselves, to appearances that exist only in representation} (B 534). The phenomenal world or the empirical world is not a thing in itself, it is the world of representation, according to Kant, generated by forms of sensibility, the human understanding, and empirical intuitions. We can neither observe the unconditioned origin of the universe nor intuit empirically its infinite past with an infinite sequence of causes and effects.

Regarding our scientific cosmology, the first antinomy of pure reason speaks of the impossibility of the big bang as the scientific origin of the world and, at the same time, imposes a limit to our intention to know the past universe. The noumenon---as an epistemological limitation for models and theories, and, in particular, as a singularity or a breakdown of general relativity in the standard model---would \enquote{appear} as a limit to observations even in bouncing models or cyclic cosmologies. In sum, Kant's philosophy may help us to conceive of a better notion of science and scientificity. At the same time, it narrows its range, especially in cosmology.        

\section{Final remarks}
A careful physicist or cosmologist should never speak of the big bang as a scientific origin for the universe. The standard model in cosmology, also known as $\Lambda$CDM, has been supported up to now by data from the primordial nucleosynthesis (a few seconds to 20 minutes after the supposed big bang) until today. In the future, if cosmological gravitational waves are detected, the inflationary period will increase the known threshold to a tiny fraction of a second after the initial singularity. But the standard model is forbidden at the big bang. On the one hand, from a physical point of view, quantities like energy density and matter pressure are unbounded at the initial singularity. Then, the big bang appears as an unconditioned origin in this context. According to Kant, the empirical world is the world of conditioned phenomena, i.e., every phenomenon studied by natural sciences is subject to a condition and provides finite data. Therefore, the big bang is not a matter of science. On the other hand, from the mathematical point of view, singularity theorems do not offer a definition of singularity.  Thus, the big bang is some sort of what Kant called the noumenon, i.e., a problematic concept which provides neither data nor concept.

From the difference between to think of an object and to cognize an object, introduced by Kant in the \textit{Critique of pure reason}, I proposed a degree of scientificity using fuzzy sets. With a continuum degree, theories and models are not conceived of as exclusively scientific or not by means of a binary reasoning. By adopting a fuzzy degree of scientificity, it is possible to estimate, for example, the degree of scientificity of cosmological models. Among options shown in this article, that is to say, the standard model, emergent cosmologies, and bouncing cosmologies, the latter class provides more thought or predicted objects (from contraction phase) and, consequently, more future data. In this sense, the fuzzy set of known objects is potentially larger for bouncing cosmologies and then their degree of scientificity may surpass its rivals.

\section*{Acknowledgments}
This study was financed in part by the Coordenação de Aperfeiçoamento de Pessoal de Nível Superior - Brasil (CAPES) - Finance Code 001.


\begin{thebibliography}{}

\bibitem[Abbott et al.(2016)]{Abbott}Abbott, B. P. \textit{et al.} (LIGO Scientific Collaboration and Virgo Collaboration). (2016). Observation of gravitational waves from a binary black hole merger. \textit{Physical Review Letters, 116}, 061102.

\bibitem[Ade et al.(2016)]{PlanckColl}Ade, P. A. R. \textit{et al.} (Planck Collaboration). (2016). Planck 2015 results XIII. Cosmological parameters. \textit{Astronomy and Astrophysics, 594}, A13.

\bibitem[Al-Azm(1968)]{Al-Azm}Al-Azm, S. J. (1968). Kant's conception of the noumenon. \textit{Dialogue, 6} (4), 516-520.

\bibitem[Anderson(1998)]{Anderson}Anderson, R. L. (1998). Truth and objectivity in perspectivism. \textit{Synthese, 115}, 1-32.

\bibitem[Ansoldi(2007)]{Ansoldi}Ansoldi, S. (2007). Spherical black holes with regular center: a review of existing models including a recent realization with Gaussian sources. In: Proceedings of  BH2, \textit{Dynamics and Thermodynamics of Black Holes and Naked Singularities}. Milano, Italy. 

\bibitem[Ashtekar et al.(2006)]{Ashtekar}Ashtekar, A., Pawlowski, T., and Singh, P. (2006). Quantum nature of the big bang: An analytical and numerical investigation. \textit{Physical Review D,73}, 124038.

\bibitem[Babich(1999)]{Babich}Babich, B. (ed.). (1999). \textit{Nietzsche, epistemology, and philosophy of science}. London: Kluwer Academic Publishers.

\bibitem[Bag et al.(2014)]{Bag}Bag, S., Sahni, V., Shtanov, Y., and Unnikrishnan, S. (2014). Emergent cosmology revisited. \textit{Journal of Cosmology and Astroparticle Physics, 07}, 034.  

\bibitem[Barrau et al.(2017)]{Barrau}Barrau, A., Martineau, K., and Moulin, F. (2017). Seeing through the cosmological bounce: Footprints of the contracting phase and luminosity distance in bouncing models. \textit{Physical Review D, 96}, 123520.

\bibitem[Borde et al.(2003)]{Borde}Borde, A., Guth, A. H., and Vilenkin, A. (2003). Inflationary spacetimes are incomplete in past directions. \textit{Physical Review Letters, 90}, 151301.

\bibitem[Brandenberger and Peter(2017)]{Brandenberger}Brandenberger, R., and Peter, P. (2017). Bouncing cosmologies: progress and problems. \textit{Foundations of Physics, 47} (6), 797-850. 

\bibitem[Carroll(2001)]{Carroll2}Carroll, S. M. (2001). The cosmological constant. \textit{Living Reviews in Relativity, 4}, 1.

\bibitem[Carroll(2004)]{Carroll}Carroll, S. (2004). \textit{Spacetime and geometry: an introduction to general relativity}. San Francisco: Addison Wesley.

\bibitem[Coc and Vangioni(2017)]{Coc}Coc, A. and Vangioni, E. (2017). Primordial nucleosynthesis. \textit{International Journal of Modern Physics E, 26} (7), 1741002.

\bibitem[Dorato(2002)]{Dorato}Dorato, M. (2002). Kant, Gödel and relativity. In: P. Gardenfors, K. Kijania-Placek and J. Wolenski (eds.), \textit{Proceedings of the invited papers for the 11th International Congress of the Logic Methodology and Philosophy of Science}, Synthese Library, Kluwer, Dordrecht, 329-346.

\bibitem[Einstein(1916)]{Einstein}Einstein, A. (1916). Die Grundlage der allgemeinen Relativitätstheorie. \textit{Annalen der Physik, 354} (7), 769-822.

\bibitem[Einstein(1917)]{Einstein2}Einstein, A. (1917). Kosmologische Betrachtungen zur allgemeinen Relativitätstheorie. \textit{Sitzungsberichte der Königlich Preu\ss ischen Akademie der Wissenschaften}, 142-152.

\bibitem[Ellis and Maartens(2004)]{Ellis}Ellis, G. F. R. and Maartens, R. (2004). The emergent universe: inflationary cosmology with no singularity. \textit{Classical and Quantum Gravity, 21}, 223-232.

\bibitem[Freese(2017)]{Freese}Freese, K. (2017). Status of dark matter in the universe. \textit{International Journal of Modern Physics D, 26}, 1730012. 

\bibitem[Friedmann(1922)]{Friedmann}Friedmann, A. (1922). Über die Krümmung des Raumes. \textit{Zeitschrift für Physik A, 10} (1) 377. 

\bibitem[Gava(2014)]{Gava}Gava, G. (2014). Kant?s definition of science in the \textit{Architectonic of pure reason}
and the essential ends of reason. \textit{Kant-Studien, 105 (3)}, 372-393.

\bibitem[Grier(2018)]{Grier}Grier, M. (2018). Kant's critique of metaphysics. \textit{The Stanford Encyclopedia of Philosophy} (Summer 2018 Edition), E. N. Zalta (ed.), URL = <https://plato.stanford.edu/ \ archives/sum2018/entries/kant-metaphysics/>.

\bibitem[Guth(1981)]{Guth}Guth, A. H. (1981). The inflationary universe: a possible solution to the horizon and flatness problems. \textit{Physical Review D, 23}, 347-356.

\bibitem[Hawking and Ellis(1973)]{Hawking}Hawking, S.W. and Ellis, G. F. R. (1973). \textit{The large scale structure of space-time}. Cambridge: Cambridge University Press.

\bibitem[Hawking and Penrose(1970)]{Hawking2}Hawking, S.W. and Penrose, R. (1970). The singularities of gravitational collapse and cosmology. \textit{Proceedings of the Royal Society of London A, 314}, 529-548.

\bibitem[Ijjas et al.(2013)]{Steinhardt}Ijjas, A., Steinhardt, P. J., and Loeb, A. (2013). Inflationary paradigm in trouble after Planck 2013. \textit{Physics Letters B, 723}, 261.

\bibitem[Ijjas et al.(2014)]{Steinhardt2}Ijjas, A., Steinhardt, P. J., and Loeb, A. (2014). Inflationary schism. \textit{Physics Letters B, 736}, 142-146.

\bibitem[Janiak(2004)]{Janiak}Janiak, A. (2004). Kant as philosopher of science. \textit{Perspectives on Science, 12 (3)},  339-363.

\bibitem[Kant(1998)]{Kant}Kant, I. (1998). \textit{Critique of pure reason}, translated by Paul Guyer and Allen W. Wood.  Cambridge: Cambridge University Press.

\bibitem[Kant(2004)]{Kant2}Kant, I. (2004). \textit{Prolegomena to any future metaphysics}, translated by Gary Hatfield. Cambridge: Cambridge University Press.

\bibitem[Leech(2017)]{Leech}Leech, J. (2017). Kant's material condition of real possibility. In M. Sinclair (ed.), \textit{The actual and the possible: modality and metaphysics in modern philosophy}. Oxford: Oxford University Press.

\bibitem[Lehners(2008)]{Lehners}Lehners, J. L. (2008). Ekpyrotic and cyclic cosmology. \textit{Physics Reports, 465}, 223.

\bibitem[Lemaître(1931)]{Lemaitre}Lemaître, G. (1931). A homogeneous universe of constant mass and increasing radius accounting for the radial velocity of extra-galactic nebulæ. \textit{Monthly Notices of the Royal Astronomical Society, 91}, 483.

\bibitem[Linde(2015)]{Linde}Linde, A. (2015). Inflationary cosmology after Planck 2013. In C. Deffayet, P. Peter, B. Wandelt, M. Zaldarriaga, and L. F. Cugliandolo (eds), \textit{Post-Planck Cosmology: Lecture Notes of the Les Houches Summer School: Volume 100, July 2013}. Oxford: Oxford University Press. 

\bibitem[Merritt(2017)]{Merritt}Merritt, D. (2017). Cosmology and convention. \textit{Studies in History and Philosophy of Modern Physics, 57}, 41-52.  

\bibitem[Naraniecki(2010)]{Naraniecki}Naraniecki, A. (2010). Neo-Positivist or Neo-Kantian? Karl Popper and the Vienna Circle. \textit{Philosophy, 85} (4), 511-530.

\bibitem[Neves(2016)]{Neves4}Neves, J. C. S. (2016). Are black holes in an ekpyrotic phase possible?. \textit{Astrophysics and Space Science,  361}, 281.

\bibitem[Neves(2017)]{Neves2}Neves, J. C. S. (2017). Bouncing cosmology inspired by regular black holes. \textit{General Relativity and Gravitation, 49}, 124.

\bibitem[Neves(2019a)]{Neves1}Neves, J.C.S. (2019a). Nietzsche for physicists. \textit{Philosophia Scienti{\ae}, 23} (1), 185-201.

\bibitem[Neves(2019b)]{Neves3}Neves, J. C. S. (2019b). Infinities as natural places. \textit{Foundations of Science, 24} (1), 39-49.

\bibitem[Neves and Saa(2014)]{Neves_Saa}Neves, J. C. S. and Saa, A. (2014). Regular rotating black holes and the weak energy condition. \textit{Physics Letters B, 734}, 44-48. 

\bibitem[Nietzsche(2002)]{Beyond}Nietzsche, F. (2002). \textit{Beyond good and evil}, translated by Judith Norman. Cambridge: Cambridge University Press. 

\bibitem[Nietzsche(2003)]{Fragments}Nietzsche, F. (2003). \textit{Writings from the late notebooks}, translated by Kate Sturge. Cambridge: Cambridge University Press. 

\bibitem[Nietzsche(2005)]{Ecce}Nietzsche, F. (2005). \textit{The Anti-Christ, Ecce Homo and Twilight of the Idols}, translated by Judith Norman.  Cambridge: Cambridge University Press. 

\bibitem[Novello and Perez Bergliaffa(2008)]{Novello}Novello, M. and Perez Bergliaffa S. E. (2008). Bouncing cosmologies. \textit{Physics Reports, 463}, 127-213.

\bibitem[Palmquist(1986)]{Palmquist}Palmquist, S. R. (1986). Six perspectives on the object in Kant's theory of knowledge. \textit{Dialectica, 40} (2), 121-151.

\bibitem[Planck(1900)]{Planck}Planck, M. (1900). Zur Theorie des Gesetzes der Energieverteilung im Normalspektrum. \textit{Verhandlungen der Deutschen Physikalischen Gesellschaft, 2,} 237.  

\bibitem[Plato(1931)]{Plato}Plato. (1931). \textit{Timaeus}, translated by B. Jowett. London: Oxford University Press.

\bibitem[Popper(1994)]{Popper3} Popper, K. (1994). \textit{Objective knowledge: an evolutionary approach}. Oxford: Clarendon Press. 

\bibitem[Popper(2005a)]{Popper}Popper, K. (2005a). \textit{The logic of scientific discovery}. London-New York: Routledge Classics.

\bibitem[Popper(2005b)]{Popper2}Popper, K. (2005b). \textit{Unended quest: an intellectual autobiography}. London-New York: Routledge Classics.

\bibitem[Pula(1992)]{Pula}Pula, R. P. (1992). The Nietzsche-Korzybski-Sapir-Whorf Hypothesis?. \textit{ETC: A Review of General Semantics, 49} (1) 50-57. 

\bibitem[Quintin and Brandenberger(2016)]{Quintin}Quintin, J., and Brandenberger, R. H. (2016). Black hole formation in a contracting universe. \textit{Journal of Cosmology and Astroparticle Physics, 11}, 029. 

\bibitem[Robertson(1935)]{Robertson}Robertson, H. P. (1935). Kinematics and world structure. \textit{Astrophysical Journal,
82}, 284.

\bibitem[Romero(2013)]{Romero}Romero, G. E. (2013). Adversus singularitates: the ontology of space-time singularities. \textit{Foundations of Science, 18} (2), 297-306. 

\bibitem[Starobinsky(1980)]{Starobinsky}Starobinsky, A. A. (1980). A new type of isotropic cosmological models without singularity. \textit{Physics Letters B, 91}, 99-102.

\bibitem[Wald(1984)]{Wald}Wald, R. M. (1984). \textit{General relativity}. Chicago: The University of Chicago Press.

\bibitem[Walker(1937)]{Walker}Walker, A. G. (1937). On Milne's theory of world-structure. \textit{Proceedings of the London Mathematical Society, 2-42} (1), 90.

\bibitem[Watkins and Stan(2014)]{Watkins}Watkins, E., and Stan, M. (2014). Kant's Philosophy of Science. \textit{The Stanford Encyclopedia of Philosophy} (Fall 2014 Edition), E. N. Zalta (ed.), URL = <https:// \ plato.stanford.edu/archives/fall2014/entries/kant-science>.

\bibitem[Weinberg(2014)]{Weinberg}Weinberg, S. (2014). \textit{Cosmology}. Oxford: Oxford University Press. 

\bibitem[Werkmeister(1977)]{Werkmeister}Werkmeister, W. H. (1977). The critique of pure reason and physics. \textit{Kant-Studien, 68}, 33-45. 

\bibitem[van den Berg(2011)]{Berg}van den Berg, H. (2011). Kant?s conception of proper science. \textit{Synthese, 183}, 7-26.

\bibitem[Zadeh(1965)]{Zadeh}Zadeh, L. A. (1965). Fuzzy sets. \textit{Information and Control, 8}, 338-353.

\bibitem[Zimmermann(1996)]{Zimmermann1}Zimmermann, H.-J. (1996). \textit{Fuzzy set theory---and its applications}. Massachusetts: Kluwer Academic Publishers.

\bibitem[Zimmermann(2010)]{Zimmermann2}Zimmermann, H.-J. (2010). Fuzzy set theory. \textit{WIREs Computational Statistics, 2}, 317-332.


\end{thebibliography}
\end{document}